\documentclass[twocolumn, switch]{article} 
\usepackage{preprint}
\usepackage{enumitem}
\usepackage{amsmath, amsthm, amssymb, amsfonts}
\usepackage{subfigure}

\usepackage[numbers,sort&compress]{natbib}
\usepackage[utf8]{inputenc}	
\usepackage[T1]{fontenc}	
\usepackage{xcolor}		
\usepackage[colorlinks = true,
            linkcolor = purple,
            urlcolor  = blue,
            citecolor = black,
            anchorcolor = black]{hyperref}	
\usepackage{booktabs} 		
\usepackage{nicefrac}		
\usepackage{microtype}		
\usepackage{lineno}		
\usepackage{float}			

\usepackage{lipsum}		

\usepackage{newfloat}
\DeclareFloatingEnvironment[name={Supplementary Figure}]{suppfigure}
\usepackage{sidecap}
\sidecaptionvpos{figure}{c}

\usepackage{titlesec}
\titlespacing\section{0pt}{12pt plus 3pt minus 3pt}{1pt plus 1pt minus 1pt}
\titlespacing\subsection{0pt}{10pt plus 3pt minus 3pt}{1pt plus 1pt minus 1pt}
\titlespacing\subsubsection{0pt}{8pt plus 3pt minus 3pt}{1pt plus 1pt minus 1pt}

\DeclareMathOperator*{\argmin}{argmin}
\DeclareMathOperator*{\argmax}{argmax}

\title{Optimal Charging Method for Effective Li-ion Battery Life Extension Based on Reinforcement Learning }

\usepackage{xwatermark}

\usepackage{authblk}

\author[1]{Minho Kim}
\author[2]{Jongchan Baek}
\author[3\thanks{soohee.han@postech.ac.kr}]{Soohee Han}
\affil[1,2,3]{Department of Creative IT Engineering, Pohang University of Science and Technology}

\backgroundsetup{contents={}}

\begin{document}

\twocolumn[ 
  \begin{@twocolumnfalse} 
  
\maketitle

\begin{abstract}
A reinforcement learning-based optimal charging strategy is proposed for Li-ion batteries to extend the battery life and to ensure the end-user convenience. Unlike most previous studies that do not reflect real-world scenario well, in this work, end users can set the charge time flexibly according to their own situation rather than reducing the charge time as much as possible; this is possible by using soft actor-critic (SAC), which is one of the state-of-the-art reinforcement learning algorithm. In this way, the battery is more likely to extend its life without disturbing the end users. The amount of aging is calculated quantitatively based on an accurate electrochemical battery model, which is directly minimized in the optimization procedure with SAC. SAC can deal with not only the flexible charge time but also varying parameters of the battery model caused by aging once the offline learning is completed, which is not the case for the previous studies; in the previous studies, time-consuming optimization has to be implemented for each battery model with a certain set of parameter values. The validation results show that the proposed method can both extend the battery life effectively and ensure the end-user convenience.
\end{abstract}
\vspace{0.35cm}

  \end{@twocolumnfalse} 
] 



\section{Introduction}
Li-ion batteries have many advantages over the other kinds of batteries in many ways. That is, they have high power and energy density and the rechargeability with no memory effect, which leads to wide application to many applications such as electric vehicles (EVs), portable devices, and large scale energy storage systems (ESSs).  However, the Li-ion batteries suffer from two major challenges related to charging: the aging of the battery and the inconvenience of end-users. According to many rules of thumb in many studies, there is a trade-off between them. Unlike several kinds of fossil fuel, Li-ion battery charging must be carried out very carefully, since the charging method greatly affects how actively electrochemical side reactions occur inside the battery. Although low C-rate charging is believed to be good for health according to the rules of thumb in the Li-ion battery community, this can be inconvenient for end-users because low C-rate means long charge time. Therefore, various optimal charging methods have been developed by many researchers to minimize charge time and extend the battery life at the same time. However, there have been many weaknesses in such previous studies.

In most previous studies, the main focus is on minimizing charging time \cite{klein2011optimal,torchio2015real,perez2016optimal, perez2017optimal, zou2017electrochemical, yin2019new, park2020reinforcement}. In other words, the objective function to directly minimize is charge time in most studies. Actually, it is not reflecting the real-world battery charging scenario; in real-world, we don't always have to charge the battery in a short time. For example, while we are sleeping, we only need to charge the vehicle until we go out with our electric vehicles (EVs) the next morning. It would be better to charge the battery slowly to extend the battery life given such long time, which leads to  both end-users' convenience and battery life extension. Tt is more practical and effective to have charge time as a constraint that can be set by the end-user rather than to minimize the charge time as much as possible. 

Most previous studies have not directly minimized quantitatively defined degree of the battery aging\cite{liu2005search,klein2011optimal,suthar2014optimal,torchio2015real,perez2016optimal,zou2017electrochemical,yin2019new,park2020reinforcement}. They just apply constraints on voltage, current, temperature, or electrochemical states to slow down the battery aging according to well-known rules of thumb in the Li-ion battery community; for example, too high C-rate, too high or low temperature, and too high or low state of charge (SOC) are believed to accelerate the battery aging. Satisfying such constraints does not mean the amount of aging is minimized though it may limit the aging to a certain level. Directly reducing the amount of aging within the charge time decided by end-users is a better solution for the end-users' convenience and the battery life extension.

Although there are studies to directly minimize quantitatively defined degree of the battery aging\cite{bashash2011plug,le2016optimal,perez2017optimal}, they do not define the amount of battery aging accurately; they define it based on simplified aging model such as the empirical models or linearized electrochemical models. In those studies, the amount of aging cannot be minimized in the correct way because the amount of aging is calculated approximately.

There was an attempt to use accurate electrochemical models to calculate the amount of battery aging and to directly minimize the aging in a given charge time. Pozzi et al.\cite{pozzi2018film} try to find an optimal charging current profile to minimize the thickness of solid electrolyte interface (SEI) layer, which is one of dominant side reaction while the battery is aging. However,in this work, obtained optimal charging current profile cannot be practically used in real world. When we try to find an optimal charging current profile for a arbitrarily given Li-ion battery, we have to identify the battery model parameters and to find an optimal charging current profile for that identified battery model, which is very time consuming. Not only the work of Pozzi et al. but also many other studies\cite{bashash2011plug,klein2011optimal,le2016optimal,liu2005search,perez2016optimal,perez2017optimal,suthar2014optimal,torchio2015real,yin2019new,zou2017electrochemical} have this crucial problem in real-world application. To the authors knowledge, only Park et al.\cite{park2020reinforcement} address this problem with a machine learning technique. However, Park et al.\cite{park2020reinforcement} just try to minimize the charge time and to apply safety constraints, which cannot well reflect real-world scenario and thus cannot minimize the amount of the battery aging effectively as mentioned above.

In this paper, a novel method to optimize the charging current for Li-ion batteries using a model-free reinforcement learning algorithm is proposed to enhance the end-user convenience and extend the Li-ion battery life at the same time. In this study, charge time can be set by the end users so that the battery life can be extended as much as possible when the end users have enough time for charging. It is noted that the proposed method can equivalently deal with fast charging problem like above-mentioned previous works because charge time can be set to any value in the available range; the available range can be automatically learned by the reinforement learning algorithm. In the proposed method, the amount of battery aging is defined quantitatively and accurately based on an electrochemical Li-ion battery model, called SPMeT\cite{moura2016battery} and optimization is conducted to directly minimize the amount of battery aging with several safety constraints. The amount of aging is defined by using the aging model suggested by Yang et al.\cite{yang2019coupled}; this aging model is combined with above-mentioned SPMeT. The optimal charging strategy is learned by using soft actor-critic (SAC)\cite{haarnoja2018soft}, which is one of the state-of-the-art model-free off-policy reinforcement learning algorithms. Unlike other previously proposed methods in the literature, SAC can come up with proper optimal current profiles according to varying end users' demands for the charge time or the state of health (SOH) of the battery once offline learning is completed. Online learning is also possible with SAC to enhance the charging strategy because of its powerful advantage that off-policy learning is possible. Hindsight experience replay (HER) \cite{andrychowicz2017hindsight} technique is also used to enhance the sample efficiency of SAC to accelerate the learning procedure. In this work, HER is modified to be suitable for the optimal charging problem. To validate the proposed method, optimal charging result of it is compared with one of constant current (CC) charging strategy. It is shown that the proposed method can effectively extend the battery life without reducing the end-user convenience.

\section{Soft actor-critic (SAC)}
Soft actor-critic (SAC) is a key algorithm in the proposed method, which is one of the state-of-the-art reinforcement learning (RL) algorithms. Problem formulation of RL is introduced and then how SAC can solve the problem is explained in this section.  

\subsubsection{Reinforcement Learning (RL)}

RL is a framework for solving a special problem that is defined based on Markov decision process (MDP). In the MDP, state and action is defined in the discrete time. State transition occurs by taking a certain action every time step. Next state $ s_{t+1} $ at time step $ t+1 $ is only dependent on the previous state $ s_{t} $ and the action $ a_{t} $ taken at time step $ t $. $ a_{t} $ is sampled from the distribution
\begin{equation} 
a_{t}\sim\pi(a_{t}|s_{t}) 
\label{policy_eq}
\end{equation}
which represents the policy for choosing the action at each time step. $ s_{t+1} $ is sampled from a certain distribution denoted by
\begin{equation}
s_{t+1}\sim p_{\text{MDP}}(s_{t+1}|s_{t},a_{t})
\label{MDP_eq}
\end{equation} 

which can explain how the state transition occurs. MDP is defined by this distribution. In real world, this distribution correspond to model equations we are dealing with. Every time the state transition occurs, reward is given. The value of reward varies according to how much desirable the state transition is. Reward $ r $ is a function of  $ s_{t} $, $ a_{t} $, and $ s_{t+1} $ as follows:
\begin{equation}
r_{t+1}=f_{\text{reward}}(s_{t},a_{t},s_{t+1})
\end{equation}
The reward function can be designed to attract the state transition of a given MDP to the desirable direction. One episode is defined to be the history of sate, action, and rewards obtained from the initial state to terminal sate; state transition stops when a predefined terminal condition is satisfied in most problems. 

\begin{figure}[h!]\centering
	\includegraphics[width=0.5\textwidth]{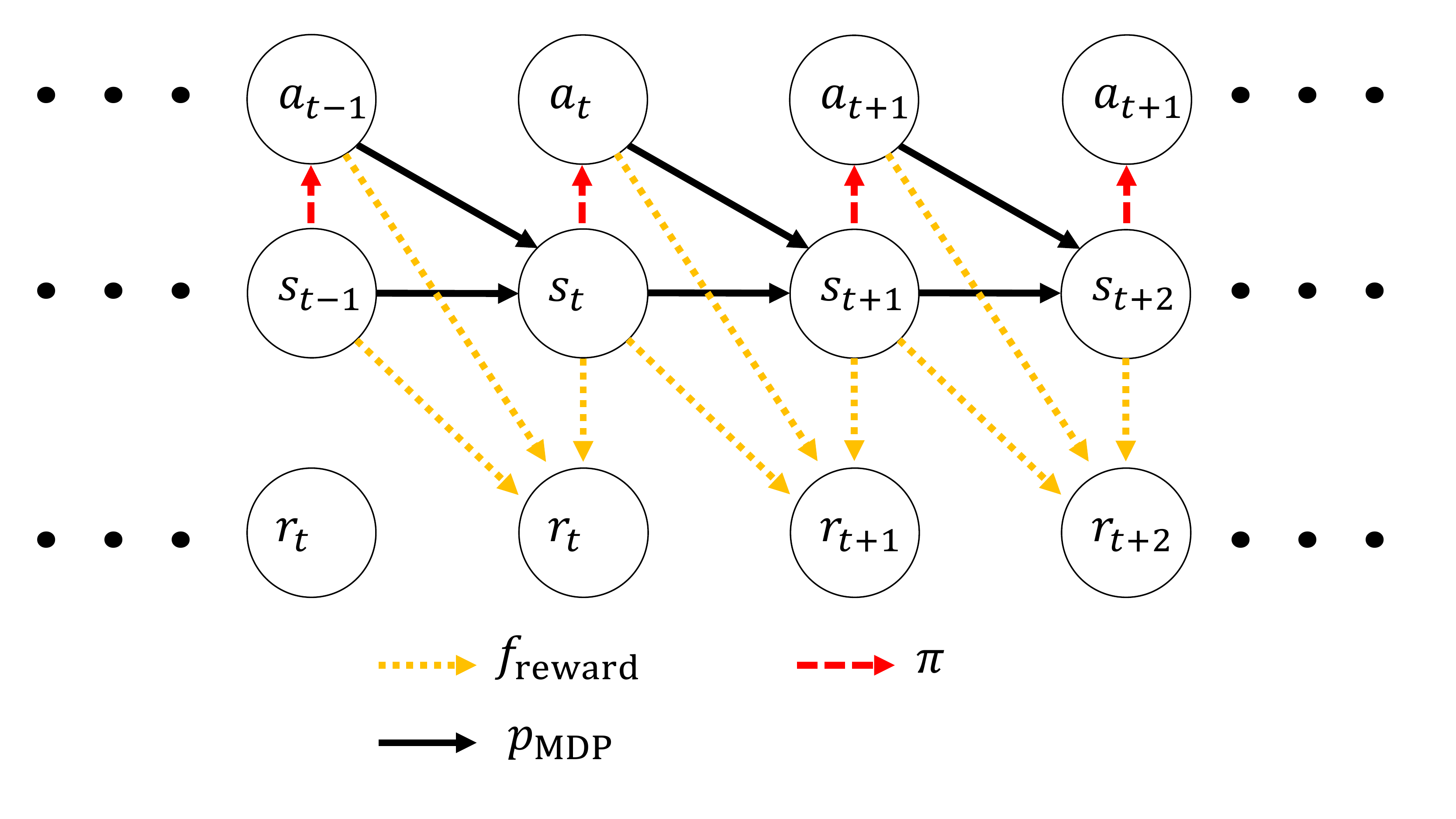}
	\caption{Markov decision process }\label{MDP_figure}
\end{figure}
\setlength{\tabcolsep}{3.1pt}
The graphical model for MDP is described in the figure. \ref{MDP_figure}.
In most cases in real world, even with the same amount of reward, it is more beneficial to receive it earlier. Therefore, the sum of rewards from the current time step $ t $ to the end of an episode is defined as follows:
\begin{equation}
R=\sum_{t=0}^{\infty}{\gamma^{t} r_{t+1}}
\label{reward_sum_eq}
\end{equation} 
where $ T $ is final time step of the episode and $ \gamma $ is discounting factor which is set to a value between $ 0 $ and $ 1 $. \textbf{The main goal of the RL is to find an optimal policy $ \pi(a|s) $ to maximize $ R $ in (\ref{reward_sum_eq}).} 
There are two important function in the RL framework: state value function and state-action value function. These functions are dependent on the policy $ \pi(a|s) $ in (\ref{policy_eq}). The state value function is defined as follows:
\begin{equation}
V^{\pi}(s)=\mathbb{E}_{s_{0}=s,\ a_t\sim\pi \ \text{for}\ t\geqq 0}[R]
\end{equation}
which is the expectation of the discounted sum of rewards $ R $ when assuming that $ s_{0}=s $ and the action is sampled from the policy distribution $ \pi $. The state-action value is defined as follows:
\begin{equation}
Q^{\pi}(s,a) =  \mathbb{E}_{s_{0}=s,\ a_{0}=a,\ a_k\sim\pi \ \text{for}\ t\geqq 1}[R]
\end{equation}
which is the expectation of the discounted sum of rewards $ R $ when assuming that $ s_{0}=s,\ a_{0}=a $ and the action is sampled from the policy distribution $ \pi $ from the time step $ t=1 $.
\textbf{The optimal policy $ \pi^* $ is the policy maximizing these two function}; actually, maximizing $ V^{\pi}(s) $ is equivalent to maximizing $ Q^{\pi}(s,a) $.
\begin{equation}
\pi^*=\argmin_{\pi}{V^{\pi}(s)}
\label{RL_problem_eq1}
\end{equation}

\begin{equation}
\pi^*=\argmin_{\pi}{Q^{\pi}(s,a)}
\label{RL_problem_eq2}
\end{equation}

\subsection{Soft actor-critic (SAC)} \label{SAC}
Soft actor-critic(SAC) is one of the sate-of-the-art off-policy model-free RL algorithm for solving the problem defined in (\ref{RL_problem_eq1}) and (\ref{RL_problem_eq2}). Off-policy means the training data do not have to be sampled from the current policy to improve the current policy. Model-free means optimal policy can be obtained without the information of the MDP. In SAC, for better convergence property and sample efficiency of the training phase, both the total rewards and the entropy of the policy distribution is maximized. The objective of SAC is to find optimal policy $ \pi $ to maximize the discounted sum of rewards and entropy as follows:
\begin{equation}
\pi^*=\argmax_{\pi}{\sum_{t=0}^{\infty}{\mathbb{E}\left[\gamma^{t} ( r_{t+1} + \mathcal{H}(\pi(\cdot|s_t)) )\right]}}
\label{reward_sum_SAC_eq}
\end{equation}
where $ \mathcal{H} $ means entropy ($ \mathcal{H}(\pi(\cdot|s_k))=\mathbb{E}_{a\sim\pi(\cdot|s_t)}[-\log(\pi(a|s_t))] $) and $ \mathbb{E} $ means expectation. 
Large entropy of the policy distribution means high randomness of action which leads to active exploration. That is, maximizing the entropy prevents the policy from falling into poor local optima caused by lack of exploration. Too much entropy is also prevented because it means low reward in the perspective of the objective function in (\ref{reward_sum_SAC_eq}). Therefore, SAC can balance exploration and exploitation, leading to sample-efficient and stable learning. Soft value function can be defined as:
\begin{equation}
V^{\pi}(s) = \sum_{t=0}^{\infty}\mathbb{E}_{s_{0}=s,\ a_t\sim\pi \ \text{for}\ t\geqq 0}\left[ \gamma^{t} ( r_{t+1} + \mathcal{H}(\pi(\cdot|s_t)) ) \right]
\label{soft_v_eq}
\end{equation}
Soft action-value function can be defined as:
\begin{equation}
Q^{\pi}(s,a) = \sum_{t=0}^{\infty}\mathbb{E}_{s_{0}=s,a_{0}=a \ a_t\sim\pi \ \text{for}\ t\geqq 1}\left[ \gamma^{t} ( r_{t+1} + \gamma\ \mathcal{H}(\pi(\cdot|s_{t+1})) ) \right]
\label{soft_q_eq}
\end{equation}
Using the definitions in (\ref{soft_v_eq}) and (\ref{soft_q_eq}), the following relation between soft state and state-action value functions:
\begin{equation}
V^{\pi}(s) = \mathbb{E}_{a\sim\pi}\left[Q^{\pi}(s,a)-\log{\pi(a|s)}\right] 
\label{soft_v_q_rel}
\end{equation}
\begin{equation}
Q^{\pi}(s,a) = \mathbb{E}_{s'\sim p_{\text{MDP}}}\left[f_{\text{reward}}(s,a,s')
+ \gamma V^{\pi}(s')\right]
\label{soft_q_v_rel}
\end{equation}
In SAC, there are two groups of neural networks: actor and critic. In actor group, there is a policy network $ \pi_{\phi}(a|s) $ parameterized with $ \phi $ which represent the policy distribution. It is noted that not only the probability density value of the policy distribution can be calculated but also sampling action from the distribution must be possible for training. In the critic group, there are network $ V_{\psi}(s) $ parameterized with $ \psi $ and network $ Q_{\theta}(s,a) $ parameterized with $ \theta $, which represent soft state value function in (\ref{soft_v_eq}) and state-action value function in (\ref{soft_q_eq}), respectively. 
\begin{figure}[h!]\centering
	\includegraphics[width=0.5\textwidth]{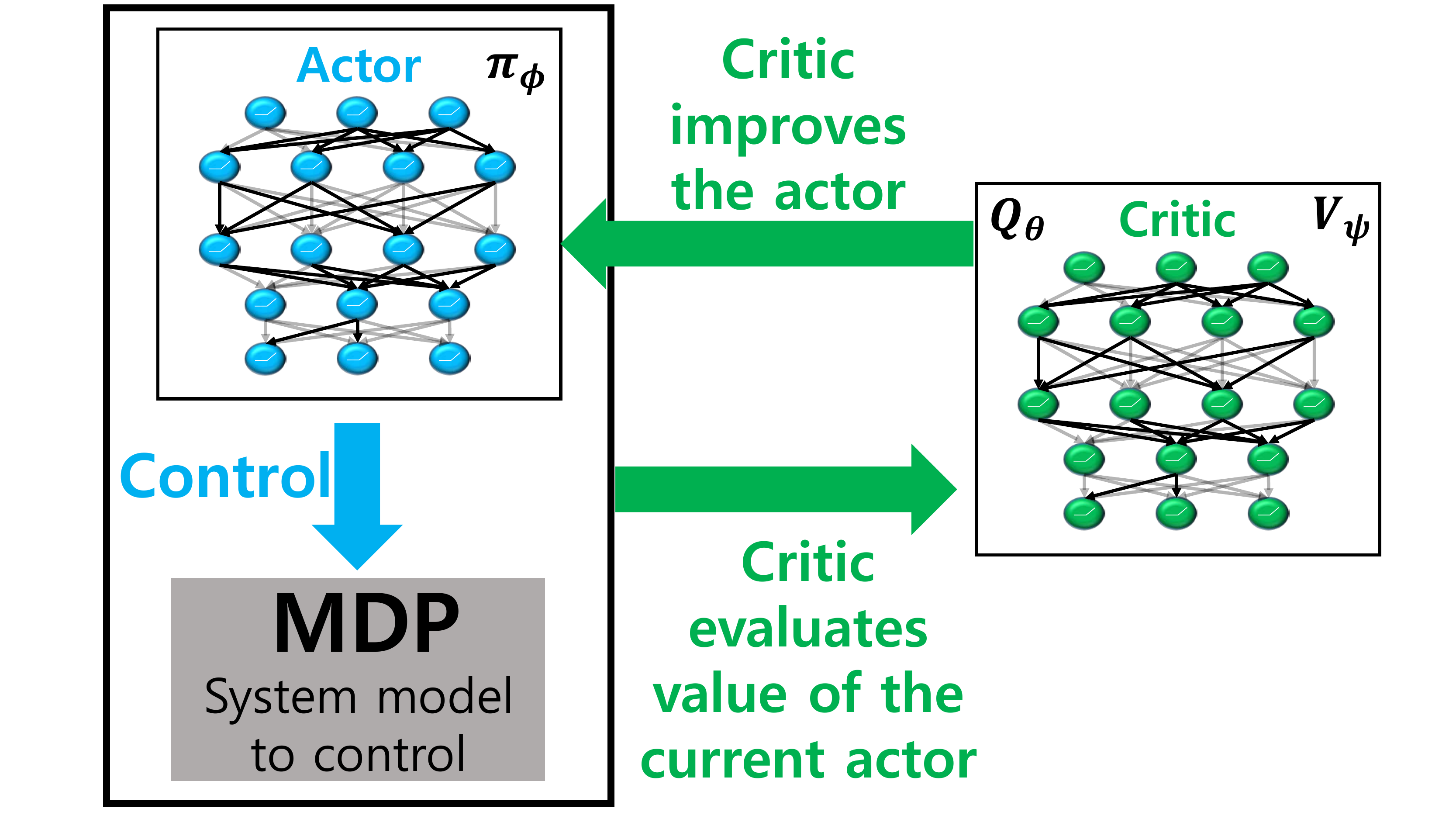}
	\caption{Concept of actor and critic }\label{actor_critic}
\end{figure}
\setlength{\tabcolsep}{3.1pt}
The role of actor and critic is described in the figure. \ref{actor_critic}. 
The actor controls the MDP. The critic observes how the current actor control the MDP and then modify the actor based on the observation to obtain bigger reward.

\textbf{The training process of SAC is as follows. At every iteration, three tasks are carried out: 1)data generation using the policy 2)value evaluation 3)policy improvement.} In data generation task, 
consecutive actions are taken according to the policy network; it means $ a_t $ is sampled from the policy network $ \pi_{\phi}(a_t|s_t)  $ every time step $ t $. Afterward, the data $ [s_t,a_t,r_{t+1},s_{t+1}] $ is saved in a replay buffer so that the saved data can be used in the training phase several times. In the value evaluation task, using the data saved in the replay buffer, $ V_{\psi}(s) $ and $ Q_{\theta}(s,a) $ are trained to represent state value and state-action value of the current policy. This is implemented by minimizing the following residual errors derived from the relation of value functions in (\ref{soft_v_q_rel}) and (\ref{soft_q_v_rel}):
\begin{eqnarray}
J_{V}(\psi)=\mathbb{E}_{s_t\sim\mathcal{D}, a_{t}\sim\pi_{\phi}}\left[ \frac{1}{2}\left(V_{\psi}(s_t)-\left[Q_{\theta}(s_t,a_t)-\log{\pi_{\phi}(a_t|s_t)}\right]\right)^2 \right]\\
J_{Q}(\theta)=\mathbb{E}_{[s_t,a_t,r_{t+1},s_{t+1}]\sim\mathcal{D}}\left[
\frac{1}{2}\left(Q_{\theta}(s_t,a_t)-\left[ r_{t+1} + \gamma\ V_{\bar{\psi}}(s_{t+1}) \right]\right)^2 \right]\label{Q_objective}
\end{eqnarray}
where $ \mathcal{D} $ represent the replay buffer. $ \bar{\psi} $ is the parameter of target value network which is an exponentially moving average of $ \psi $. Such target value network is employed to stabilize the training process. After such value evaluation procedure, the current policy is improved, which means $ Q^{\pi_{\text{new}}}(s,a) \geqq Q^{\pi_{\text{old}}}(s,a)$, by minimizing the following objective function:
\begin{equation}
J_{\pi}(\phi) = \mathbb{E}_{s_t\sim\mathcal{D},\varepsilon_t\sim\mathcal{N}(0,I)}\left[\log{\pi_{\phi}(f_{\phi}(\varepsilon_t;s_t)|s_t)}-Q_{\theta}(s_t,f_{\phi}(\varepsilon_t;s_t))\right]
\label{actor_obj_eq}
\end{equation}
where $ f_{\phi}(\epsilon_t;s_t) $ is $ a_t $ which is determined by random variable $ \varepsilon_t $ that is considered as constant when calculating gradient of the objective function in (\ref{actor_obj_eq}). The details about SAC is described in \cite{haarnoja2018soft}.

\section{Battery aging model}\label{battery_aging_model}

The battery aging model used in this work is based on single particle model with electrolyte and thermal dynamics (SPMeT) \cite{moura2016battery,perez2016optimal}. SPMeT is basically not able to reflect the side reactions related to the battery aging. To reflect such side reactions, one of the aging models for the solid electrolyte interface (SEI) layer growth and Li-plating \cite{yang2019coupled} is combined with the SPMeT in this work.

\subsection{SPMeT}
\begin{figure}[h!]\centering
	\includegraphics[width=0.5\textwidth]{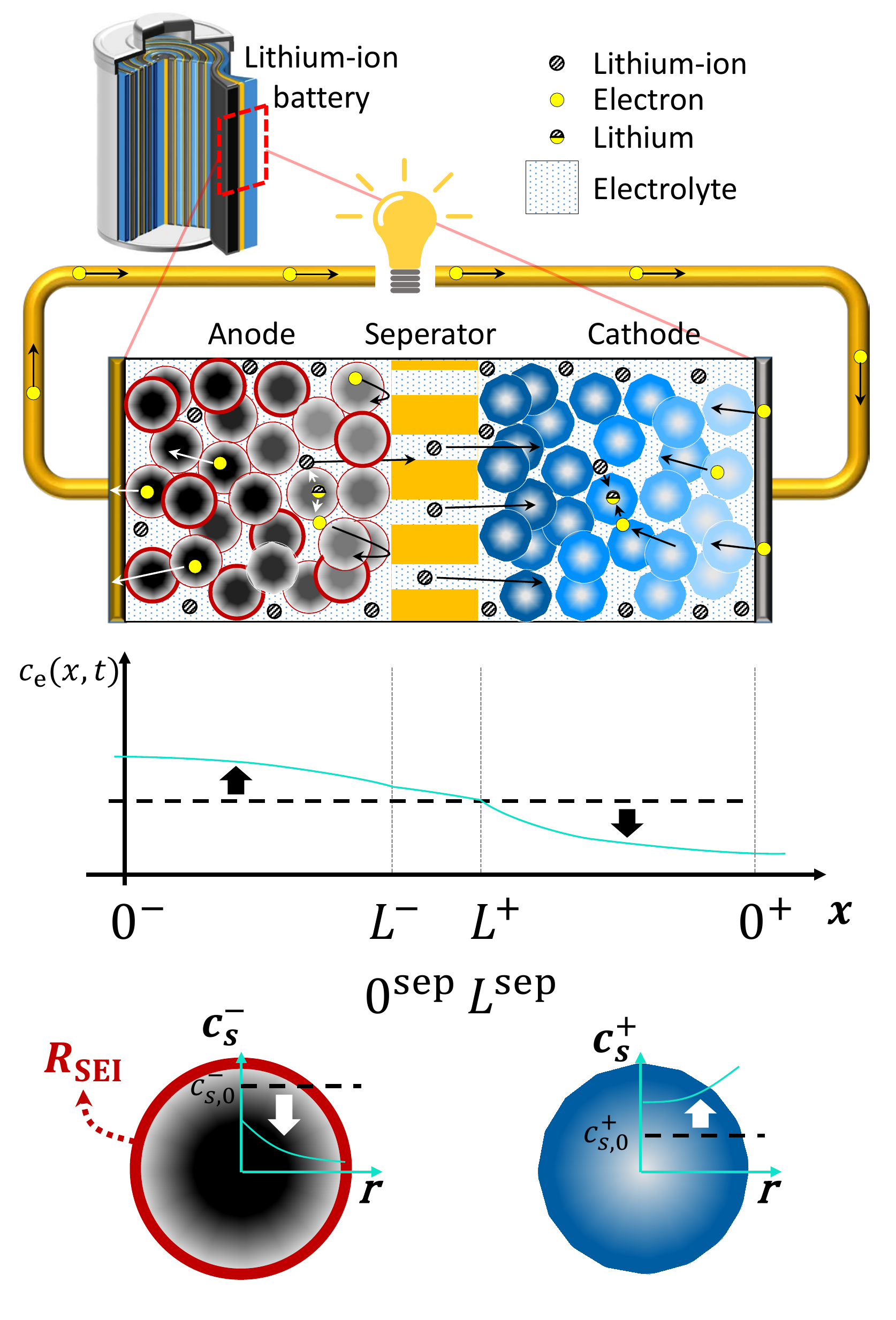}
	\caption{Schematic representation of discharge process. }\label{battery_principle}
\end{figure}

The main elements of Li-ion battery are anode, cathode, and separator. Anode and cathode have porous structure consisting of two different phase: liquid phase (electrolyte in pores) and solid phase (active material particles). The separator located between the two electrodes is also porous solid material of which liquid electrolyte infiltrates into the pores. The separator prevents internal short between two electrodes. While charging, Li-ions and electrons are transported from the cathode solid particles to the anode solid particles through the electrolyte and external circuit respectively. While discharging, they move in the opposite direction, which is described in figure \ref{battery_principle}.

\setlength{\tabcolsep}{3.1pt}

The SPMeT, which is based on the Doyle-Fuller-Newman (DFN) model\cite{doyle1993modeling,fuller1994simulation}, consists of partial differential equations (PDEs) reflecting such physical phenomena inside the Li-ion battery. The SPMeT is developed by assuming solid-phase Li concentration of each electrode is not dependent on the $ x $ coordinate (figure \ref{battery_principle}) in DFN model. Despite of such simplification, SPMeT is validated to be fairly accurate compared to DFN model \cite{moura2016battery} because it reflects key features of Li-ion battery mechanisms such as electrolyte and thermal dynamics very well. The following equations are related to the diffusion of Li in the solid particles of each electrode:
\begin{eqnarray}
\frac{\partial c_{s}^{\pm}}{\partial t}(r,t) = \frac{1}{r^{2}}\frac{\partial}{\partial r}\left[ D_{s}^{\pm}r^{2}\frac{\partial c_s^{\pm}}{\partial r}(r,t) \right],\label{eq_17}\\
\frac{\partial c_{s}^{\pm}}{\partial r}(0,t) = 0,\\
\frac{\partial c_{s}^{\pm}}{\partial r}(R_s^{\pm},t) = \mp \frac{1}{D_s^{\pm}Fa^{\pm}L^{\pm}}I(t),
\end{eqnarray}
where $ c_{s}^{\pm} $ is Li concentration in the solid phase of cathode and anode, $ I(t) $ is current of the battery per unit cross section area, which is positive for charge, $ R_s^{\pm} $ is the radius of solid particle in the cathode and anode, and $ D_s^{\pm} $ is diffusion coefficient of Li in the solid phase of cathode and anode. The equations about the dynamics of Li in the electrolyte are as follows:
\begin{eqnarray}
\frac{\partial c_e^{\pm}}{\partial t}(x,t) &=& \frac{\partial}{\partial x}\left[ D_e(c_e^{\pm})\frac{\partial c_e^{\pm}}{\partial x}(x,t) \right]\pm\frac{(1-t_c^0)}{\varepsilon_e^{\pm}FL^{\pm}}I(t),\\
\frac{\partial c_e^{\text{sep}}}{\partial t}(x,t) &=& \frac{\partial}{\partial x}\left[ D_e(c_e^{\text{sep}})\frac{\partial c_e^{\text{sep}}}{\partial x}(x,t) \right],\\
\frac{\partial c_e^{\pm}}{\partial x}(0^{\pm},t)&=&0,
\end{eqnarray}
\begin{eqnarray}
\varepsilon_e^-D_e(L^-)\frac{\partial c_e^-}{\partial x}(L^-,t)&=&\varepsilon_e^{\text{sep}}D_e(0^{\text{sep}})\frac{\partial c_e^{\text{sep}}}{\partial x}(0^{\text{sep}},t),\\
\varepsilon_e^+D_e(L^+)\frac{\partial c_e^+}{\partial x}(L^+,t)&=&\varepsilon_e^{\text{sep}}D_e(L^{\text{sep}})\frac{\partial c_e^{\text{sep}}}{\partial x}(L^{\text{sep}},t),\\
c_e(L^-,t))&=&c_e(0^{\text{sep}},t),\\
c_e(L^{\text{sep}},t))&=&c_e(L^{+},t),
\end{eqnarray}
where $ c_e^{\pm} $ is Li concentration in the electrolyte of each electrode, $ D_e $ is diffusion coefficient of Li in the electrolyte, $ \varepsilon_e^j (j\in\{+,-,\text{sep}\}) $ is porosity of each electrode and separator. Output voltage is expressed as follows:
\begin{eqnarray}
V(t)&=&\frac{RT_{\text{jel}}(t)}{\alpha F}\sinh^{-1}{\left( \frac{I(t)}{2a^+L^+\bar{i_0^+}(t)} \right)} \nonumber \\
&-&\frac{RT_{\text{jel}}(t)}{\alpha F}\sinh^{-1}{\left( \frac{-I(t)}{2a^-L^-\bar{i_0^-}(t)} \right)} \nonumber \\
&+& U^+(c_{ss}^+(t)) - U^-(c_{ss}^-(t)) \nonumber \\
&+& \left( \frac{R_f^+}{a^+L^+} + \frac{R_f^-}{a^-L^-} \right)I(t) - \left(\frac{L^+}{2\kappa^+_{\text{eff}}}+\frac{L^{\text{sep}}}{\kappa^{\text{sep}}_{\text{eff}}}+\frac{L^-}{2\kappa^-_{\text{eff}}}\right)I(t) \nonumber \\
&+& \nu(\bar{c^{+}_e}(t),T_{\text{jel}}(t))[\ln{c_e(0^+,t))}-\ln{c_e(L^+,t)}] \nonumber \\
&+& \nu(\bar{c^{\text{sep}}_e}(t),T_{\text{jel}}(t))[\ln{c_e(L^{\text{sep}},t))}-\ln{c_e(0^{\text{sep}},t)}] \nonumber \\
&+& \nu(\bar{c^{-}_e}(t),T_{\text{jel}}(t))[\ln{c_e(L^-,t))}-\ln{c_e(0^-,t)}] 
\end{eqnarray}
where $ \alpha $ is charge transfer coefficient $ [\cdot] $, $ a^{\pm} $ is specific interfacial surface area $ [\text{m}^{-1}] $, $ \bar{i_0^{\pm}}(t) = k^{\pm}(c^{\pm}_{ss}(t))^{\alpha_c} (\bar{c^{\pm}_e}(x,t)(c^{\pm}_{s,\text{max}}-c^{\pm}_{ss}(t)))^{\alpha_a} $  is the spatially averaged exchange current density $ [\text{A}/\text{m}^2] $, where $ k^{\pm} $ is kinetic reaction rate, and $ \alpha_a $ and $ \alpha_c $ are the charge transfer coefficients of oxidization and reduction respectively for the intercalation reaction, $ c_{ss}^{\pm}(t) = c_s^{\pm}(R_s^{\pm},t)$ is the surface concentration in the solid phase, $ R_f^{\pm} $ is solid-electrolyte interphase film resistance  $ [\Omega \cdot {\text{m}}^2 ] $, $ \nu(c,T)=\frac{2RT}{F}(1-t_c^0)\left( 1 + \frac{d \ln{f_{c/a}}}{d \ln{c}} \right)(c,T) $, and $ \kappa^j_{\text{eff}}  [1/\Omega\cdot \text{m}]  (j\in \{ +,-,\text{sep} \}) $ is effective conductivity of electrolyte, which is written as $ \kappa^j_{\text{eff}}=\kappa(\bar{c^j_{e}}(t)){\varepsilon_e^{j}}^{\text{Brugg}} $, where $ \bar{c^j_{e}}(t) $ is spatially averaged Li-ion concentration of the electrolyte at each electrode and separator, $ \kappa(\cdot) $ is the conductivity function of Li-ion concentration at the electrolyte, and $ \text{Brugg} $ the Bruggeman exponent. The  Lumped thermal dynamics of the battery can be obtained by defining three bulk temperature variables of jelly-roll and can of the battery and the ambient air, which are denoted by $ T_{\text{jel}} $,$ T_{\text{can}} $, and $ T_{\text{amb}} $. The heat generation rate of the jelly-roll  per unit cross section area $ Q_{\text{jel},\text{gen}} [\text{W}/\text{m}^2]$  is the sum of ohmic and entropic heat generation, which is expressed as follows:
\begin{eqnarray}
Q_{\text{jel},\text{gen}} &=& I(t)[V(t)-(U^+(\bar{c_{s}^+}(t)) - U^-(\bar{c_{s}^-}(t)))]  \nonumber \\
&&+ I(t)T_{\text{jel}}(t) \frac{\partial}{\partial T} [U^+(\bar{c_{s}^+}(t)) - U^-(\bar{c_{s}^-}(t))], \label{jel1}
\end{eqnarray}    
where $ \bar{c_{s}^{\pm}}(t) $  is the spatially averaged Li concentration in the solid particle, $ \rho_{\text{jel}} $ is the density of jelly-roll, and $ c_{p,\text{jel}} $ is the specific heat of the jelly-roll. The heat exchange rate per unit cross section area due to conduction from the can to the jelly-roll $ Q_{\text{jel},\text{cond}} [\text{W}/\text{m}^2] $ is expressed as follows:
\begin{eqnarray}
Q_{\text{jel},\text{cond}} = h_{\text{jel},\text{can}} (T_{\text{can}}-T_{\text{jel}}), \label{jel2}
\end{eqnarray}
where $ h_{\text{jel},\text{can}} [\text{W}/\text{m}^2\cdot \text{K}]$ is the heat transfer coefficient between the jelly-roll and the can. Combining (\ref{jel1}) and (\ref{jel2}), energy balance equation of jelly-roll can be derived as follows:
\begin{eqnarray}
\rho_{\text{jel}}c_{p,\text{jel}}\frac{dT_{\text{jel}}}{dt} &=&  Q_{\text{jel},\text{gen}} + Q_{\text{jel},\text{cond}}  \nonumber \\
&=&  I(t)[V(t)-(U^+(\bar{c_{s}^+}(t)) - U^-(\bar{c_{s}^-}(t)))]  \nonumber \\
&&+ I(t)T_{\text{jel}}(t) \frac{\partial}{\partial T} [U^+(\bar{c_{s}^+}(t)) - U^-(\bar{c_{s}^-}(t))] \nonumber \\
&&+ h_{\text{jel},\text{can}} (T_{\text{can}}-T_{\text{jel}}),
\end{eqnarray}
where $ \rho_{\text{jel}} [\text{kg}/\text{m}^2] $ is mass of jelly-roll per unit cross section area and $ c_{p,\text{jel}} $ is the specific heat of the jelly-roll. In a similar way, the energy balance equation for the can is derived as follows:
\begin{eqnarray}
\rho_{\text{can}}c_{p,\text{can}}\frac{dT_{\text{can}}}{dt} &=&  h_{\text{jel},\text{can}} (T_{\text{jel}}-T_{\text{can}})  \nonumber \\
&& + h_{\text{can},\text{amb}} (T_{\text{amb}}-T_{\text{can}}), 
\end{eqnarray}
where  $ \rho_{\text{can}} [\text{kg}/\text{m}^2] $ is mass of can per unit cross section area and $ c_{p,\text{can}} $ is the specific heat of the can.

\subsection{Aging model}
In this work, we have additional equations for solid electrolyte interface (SEI) layer formation and the Li-plating based on the work of Yang et al.\cite{yang2019coupled} to reflect aging effect of the battery. 
The aging model in the work of Yang et al.\cite{yang2019coupled} is simplified to make it suitable for the SPMeT.
The molar flux of SEI layer formation $ J_{\text{SEI}} [\text{mol}/\text{m}^2\cdot \text{sec}]$ at the anode particle surface can be expressed with a form of Tafel equation as follows:
\begin{eqnarray}
J_{\text{SEI}} = -\frac{c_{\text{EC},0}}{\frac{\delta_{\text{film}}}{D_{\text{EC}}}+\frac{1}{k_{\text{SEI}} \exp{\left( -\frac{\alpha_{c,\text{SEI}}F}{RT}\eta_{\text{SEI}} \right)}}}, 
\end{eqnarray}
where $ c_{\text{EC},0} $ is the concentration of ethylene carbonate (EC) in the electrolyte outside the SEI layer, $ \delta_{\text{film}} $ is the total thickness of anode particle surface layer including plated Li and SEI layer, $ D_{\text{EC}} $ is the diffusion coefficient of EC in the SEI film, $ k_{\text{SEI}} $ is	the reaction rate constant for the SEI layer formation, and $ \eta_{\text{SEI}} $ is the overpotential of the SEI layer formation reaction which is expressed as:
\begin{eqnarray}
\eta_{\text{SEI}} = \eta^{-}_{\text{int}} + U^{-}(c^{-}_{ss}(t)) - U_{\text{SEI}},
\end{eqnarray}
where $ \eta^{-}_{\text{int}} = \frac{RT_{\text{jel}}(t)}{\alpha F}\sinh^{-1}{\left( \frac{-I(t)}{2a^-L^-\bar{i_0^-}(t)} \right)} $ is the overpotential of intercalation reaction, and $ U_{\text{SEI}} $ is the equilibrium potential of the SEI layer formation. The molar flux of Li-plating  $ J_{\text{LP}} [\text{mol}/\text{m}^2\cdot \text{sec}]$ at the anode particle surface can be expressed based on the Butler-Volmer equation as follows:
\begin{flalign}
&J_{\text{LP}}  = \nonumber \\
&\min{\left\{ 0, k_{\text{LP}} \left( \frac{\bar{c^{-}_e}(t)}{c_{e,\text{ref}}} \right)^{\alpha_{a,\text{LP}}} \left[ \exp{\left( \frac{\alpha_{a,\text{LP}}F}{RT_{\text{jel}}}\eta_{\text{LP}} \right)}-\exp{\left( -\frac{\alpha_{c,\text{LP}}F}{RT_{\text{jel}}}\eta_{\text{LP}} \right)}  \right] \right\}},
\end{flalign}
where $ k_{\text{LP}} $ is the reaction rate constant for the Li-plating, $ \bar{c^{-}_e}(t) $ is the spatially averaged Li concentration of the electrolyte in the anode, $ c_{e,\text{ref}} $ is the reference Li-ion concentration in the electrolyte, which is 1 $ [\text{mol}/\text{m}^3] $, $ \alpha_{a,\text{LP}} $ and $ \alpha_{c,\text{LP}} $ are the charge transfer coefficients for the Li-plating of oxidization and reduction respectively, and  $ \eta_{\text{LP}} $ is the overpotential for the Li-plating, which is expressed as:
\begin{eqnarray}
\eta_{\text{LP}} = \eta^{-}_{\text{int}} + U^{-}(c^{-}_{ss}(t)) \label{eq_35}
\end{eqnarray} 
Arrhenius law is applied to some parameters such as $ D_s^{\pm} $, $ \kappa_{\text{eff}}^{\pm} $, $ k^{\pm} $, $ k_{\text{SEI}} $, $ D_{\text{EC}} $, and  $ k_{\text{LP}} $, which means parameter $ X $ is substituted with $ X \exp{\left( \frac{E_{X}}{R}\left( \frac{1}{T_{\text{ref}}}-\frac{1}{T} \right) \right)} $ in all the equations above, where $ E_{X} [\text{J}/\text{mol}]$ is activation energy for the parameter $ X $.

The PDEs (\ref{eq_17})-(\ref{eq_35}) of SPMeT-based battery aging model, which is defined in the continuous time and space, can be expressed as follows: 
\begin{eqnarray}
\frac{d}{dt}x(t) = f_{\text{SPMeT}}(x(t),u(t)),\\
y(t) = g_{\text{SPMeT}}(x(t),u(t)),
\end{eqnarray}
where $ x(t) $ is infinite-dimensional state of the system, $ u(t) $ is the system input which is defined as instantaneous current density $ u(t) =  I(t) $ in this system, and $ y(t) $ is the system output $ y(t)=[V(t), SOC(t), J_{\text{SEI}}(t), J_{\text{LP}}(t), T_{\text{jel}}(t), T_{\text{can}}(t)] $. The system is simulated using a numerical method with the discrete domain because there is no analytical solution for it. Therefore, from the practical point of view of solving the PDEs, $ x(t) $ is defined to be the finite number of variables in such spatially discretized domain. In real world, control of input cannot be carried out continuously, so input is assumed to be constant during each time interval $ [k\Delta t, (k+1)\Delta t] $ where $ k $ is time step index and $ \Delta t $ is time step size for control. A discrete-time system equation with the perspective of control can be formulated as follows:
\begin{eqnarray}
x_{k} &=& f^d_{\text{SPMeT}}(x_{k-1},u_{k-1}),\\
y_{k} &=& g^d_{\text{SPMeT}}(x_{k},u_{k}),
\end{eqnarray}
where $ y_{k} = [V_k, \text{SOC}_k, J_{\text{SEI},k}, J_{\text{LP},k}, T_{\text{jel},k}, T_{\text{can},k}] $ where $ V_k=V(k\Delta t) $, $ \text{SOC}_k=\text{SOC}(k\Delta t) $, $ J_{\text{SEI},k} = \int_{k\Delta t}^{(k+1)\Delta t}J_{\text{SEI}}(t)dt $ , $ J_{\text{LP},k} = \int_{k\Delta t}^{(k+1)\Delta t}J_{\text{LP}}(t)dt $, $ T_{\text{jel},k}=T_{\text{jel}}(k\Delta t) $, and $ T_{\text{can},k}=T_{\text{can}}(k\Delta t) $. $ x_k $ can be used as the MDP state $ s^{\text{MDP}}_{k} $ and $ u_k $ can be used as the MDP state $ a^{\text{MDP}}_{k} $ for the battery model so the following expression similar to (\ref{MDP_eq}) for the battery model MDP.
\begin{eqnarray}
s^{\text{MDP}}_{t+1}\sim p_{\text{SPMeT}}(s^{\text{MDP}}_{t+1}|s^{\text{MDP}}_t,a^{\text{MDP}}_t) \label{SPMeT_MDP}
\end{eqnarray}

\section{Problem formulation}
\subsection{Optimal charging problem}\label{optimal_charging_problem}
In this work, the optimal charging strategy is defined as one that minimize the SEI layer formation for the charge time allowed by the end-users while satisfying safety constraints at the same time. The safety constraints include preventing Li-plating and limiting the ranges of battery temperature, current, voltage. Such optimal charging problem can be written as:
\begin{eqnarray}
\min_{I(t)}{\int_{0}^{t_{\text{given}}}{J_{\text{SEI}}(t)dt }} \label{objective}
\end{eqnarray}
subject to the following initial state constraints:
\begin{eqnarray}
&\text{Equilibrium state with initial SOC condition}& \nonumber\\
&\text{SOC}(0)= \text{SOC}_{0},&
\end{eqnarray}
the following final state condition:
\begin{eqnarray}
&\text{SOC}(t_{\text{given}})= \text{SOC}_{\text{given}},&\label{SOC_arrival_cond}
\end{eqnarray}
and the following safety condition:
\begin{eqnarray}
&I_{\text{min}}\leq I(t) \leq I_{\text{max}}&\label{safety_1}\\
&V_{\text{min}}\leq V(t) \leq V_{\text{max}}&\label{safety_2}\\
&T_{\text{min}}\leq T_{\text{jel}}(t) \leq T_{\text{max}}\label{safety_3}&\\
&J_{\text{LP}}(t) = 0,\label{safety_4}&
\end{eqnarray}
where $ t_{\text{given}} $ is the maximum charge time allowed by the end-users, $ \text{SOC}_{\text{given}} $ is the the state of charge (SOC) level to which the end-users want the battery to be charged, and  $ [I_{\text{min}},I_{\text{max}}] $,$ [V_{\text{min}},V_{\text{max}}] $, and $ [T_{\text{min}},T_{\text{max}}] $ is allowed safe range of the battery current density, voltage, and temperature, respectively. It is noted that the SOC of the battery is defined by its open circuit voltage (OCV). That is, for given OCV vlaues, $ \text{OCV}_{\text{SOC}=0 \%} $ and $ \text{OCV}_{\text{SOC}=100 \%} $ corresponding to SOC 0 \% and 100 \% respectively, SOC is defined as follows:
\begin{align}
\text{SOC}(t) = \frac{c_{s}^{-}(t) - c_{s,\text{SOC}=0\%}^{-}}{c_{s,\text{SOC}=100\%}^{-}-c_{s,\text{SOC}=0\%}^{-}} = \frac{c_{s, \text{SOC}=0\%}^{+}-c_{s}^{+}(t) }{c_{s,\text{SOC}=0\%}^{+}-c_{s,\text{SOC}=100\%}^{+}},
\end{align} 
where 
\begin{eqnarray}
&U^{+}(c_{s, \text{SOC}=0\%}^{+}) - U^{-}(c_{s, \text{SOC}=0\%}^{-}) =  \text{OCV}_{\text{SOC}=0 \%},\\
&U^{+}(c_{s, \text{SOC}=100\%}^{+}) - U^{-}(c_{s, \text{SOC}=100\%}^{-}) =  \text{OCV}_{\text{SOC}=100 \%}
\end{eqnarray}
It is noted that unlike previous works minimizing the charge time, charge time is given by the end-users and dealt with as a time constraint to further extend the battery life. That is, by allowing long charge time when the end-users have enough time until next use of the battery, we can give more freedom to the battery operation so that the battery has more chance to delay its aging process. 

\subsection{RL problem formulation of the optimal charging problem}

Time interval between two states of MDP is set $ \Delta t = 5\ \text{sec} $ in this work to convert the continuous-time battery aging model (Section\ref{battery_aging_model}) into discrete-time MDP. To solve the optimal charging problem (Section \ref{optimal_charging_problem}) by using RL, a state, action, reward function, and the termination condition are defined as follows:
\begin{itemize}[leftmargin=*]
\item \textbf{State}: state of $ k $-th time step ($ t=k\Delta t $) is defined as a vector 
\begin{align}
\mathbf{s}_k = [t_{\text{given}}-k\Delta t, \text{SOC}_{\text{given}}, I_k^{\text{time}}, V_k^{\text{time}}, T_k^{\text{time}}, \text{SOC}(k\Delta t)]^T, 
\end{align}
where 
\begin{align}
I_k^{\text{time}} &=& [I((k-l+1)\Delta t), I((k+1)\Delta t), ..., I(k\Delta t)]&  \\
V_k^{\text{time}} &=& [V((k-l+1)\Delta t), V((k+1)\Delta t), ..., V(k\Delta t)]&  \\
T_k^{\text{time}} &=& [T_{\text{can}}((k-l+1)\Delta t), T_{\text{can}}((k+1)\Delta t), ..., T_{\text{jel}}(k\Delta t)]&  
\end{align}
where $ l $ is the size of time window of the time series history used as the state. Too large $ l $ makes learning slow and too small $ l $ makes converged policy For negative $ t $, $ I(t) $, $ V(t) $, and $ T_{text{jel}}(t) $ are defined to be the same as $ I(0) $, $ V(0) $, and $ T_{text{jel}}(0) $, respectively. For practical application of the proposed optimal charging strategy, the state is made to consist of readily measurable or easily predictable variables. This state is not exactly the same as the MDP state in (\ref{SPMeT_MDP}). However, by including time-series history of the measurable data, it is assumed that the state $ \mathbf{s}_k $ contains enough amount of information on $ s^{\text{MDP}}_k $, which means there is almost one-to-one relationship between $ \mathbf{s}_k $ and $ s^{\text{MDP}}_k $ and $ s^{\text{MDP}}_k=g(\mathbf{s}_k) $ for a certain function $ g(\cdot) $. Such function $ g(\cdot) $ is automatically obtained because the actor and critic networks are trained to find optimal features in the hidden layers. Therefore, if the time-series history is long enough, using $ \mathbf{s}_k $ as a state have almost the same effect as using $ s^{\text{MDP}}_k $.

\item \textbf{Action} : action is defined to be the normalized current density. The action of $ k $-th time step is written as follows:
\begin{eqnarray}
\mathbf{a}_k = \frac{I(k\Delta t)-I_{\text{min}}}{I_{\text{max}} - I_{\text{min}}}
\end{eqnarray}
It is noted that current density is constant while each interval $ [(k-1)\Delta t, k\Delta t] $.

\item \textbf{Termination condition}: One episode terminates only when any of the following conditions are met:
\begin{eqnarray}
t_{\text{given}}&\leq& k\Delta t \label{terminal_1}\\
\frac{I_{\text{max}}}{I_{\text{1Crate}}}\frac{(t_{\text{given}}-k\Delta t)}{3600} &\leq& \text{SOC}_{\text{given}}-\text{SOC}(k\Delta t) \label{terminal_2} \\
\frac{I_{\text{min}}}{I_{\text{1Crate}}}\frac{(t_{\text{given}}-k\Delta t)}{3600} &\geq& \text{SOC}_{\text{given}}-\text{SOC}(k\Delta t) \label{terminal_3}
\end{eqnarray}
where $ I_{\text{1Crate}} $ is current density corresponding to 1 C-rate, which is defined as:
\begin{eqnarray}
I_{\text{1Crate}} &=& \frac{\varepsilon_e^- L^- (c_{s,\text{SOC=100\%}}^{-}-c_{s,\text{SOC=0\%}}^{-})}{3600} \nonumber \\
&=& \frac{\varepsilon_e^+ (0^+-L^+) (c_{s,\text{SOC=0\%}}^{+}-c_{s,\text{SOC=100\%}}^{+})}{3600} 
\end{eqnarray}

\item \textbf{Reward function}: reward is given only when transition to terminal state occurs. Reward function is dependent to which termination condition triggers the episode termination as follows:\\
\begin{itemize}
	\item[-] Termination condition (\ref{terminal_1}): 
	\begin{eqnarray}
	\mathbf{r}_{k} =
	\begin{cases}
	 \omega_{\text{SEI}}\left( \sum_{j=0}^{k_{\text{ter}}}{J_{\text{SEI},j}} \right) \\
	 + \omega_{\text{SAF}}\left(-\sum_{j=0}^{k_{\text{ter}}}{\mathbb{I}_j^{\text{SAF}}}\right),& \text{if $ k=k_{\text{ter}} $}\\\\
	 0,& otherwise
	\end{cases}
	\end{eqnarray}
	where $ \mathbb{I}_j^{\text{SAF}} $ is $ 1 $ when the safety constraints (\ref{safety_1})-(\ref{safety_4}) are not satisfied during the time interval $ [k\Delta t, (k+1)\Delta t] $, which is otherwise $ 0 $, $ \omega_{\text{SEI}} $ and $ \omega_{\text{SAF}} $ are parameters of the reward function determining how much penalty is given to the SEI layer formation and the violation of the safety constraints, respectively, which needs to be tuned for the desirable result.

	\item[-] Termination condition (\ref{terminal_2}): 
	\begin{eqnarray}
	\mathbf{r}_{k} =
	\begin{cases}
	\omega_{\text{SEI}}\left( \sum_{j=0}^{k_{\text{ter}}}{J_{\text{SEI},j}} \right. \\
	\left. + \int_{0}^{n_{\text{max}}\Delta t}{J_{\text{SEI}}(t)dt \rvert_{x(0)=x_{k_\text{ter}},I(t)=I_{\text{max}} }} \right) \\
	+ \omega_{\text{SAF}}\left(-\sum_{j=0}^{k_{\text{ter}}}{\mathbb{I}_j^{\text{SAF}}} \right.\\
	\left. - \sum_{j=0}^{n_{\text{max}}}{\mathbb{I}_j^{\text{SAF}}  \rvert_{x(0)=x_{k_\text{ter}},I(t)=I_{\text{max}} }} \right),& \text{if $ k=k_{\text{ter}} $}\\\\
	0,& otherwise
	\end{cases}
	\end{eqnarray}
	where $ n_{\text{max}}=\left[\frac{3600(\text{SOC}_{\text{given}}-\text{SOC}_{k_{\text{ter}}})I_{\text{1Crate}}}{I_{\text{max}}t} \right]$ ($ [x] $ is maximum integer not bigger than x).
		
	\item[-] Termination condition (\ref{terminal_3}): 
	\begin{eqnarray}
	\mathbf{r}_{k} =
	\begin{cases}
	\omega_{\text{SEI}}\left( \sum_{j=0}^{k_{\text{ter}}}{J_{\text{SEI},j}} \right. \\
	\left. + \int_{0}^{n_{\text{min}}\Delta t}{J_{\text{SEI}}(t)dt \rvert_{x(0)=x_{k_\text{ter}},I(t)=I_{\text{min}} }} \right) \\
	+ \omega_{\text{SAF}}\left(-\sum_{j=0}^{k_{\text{ter}}}{\mathbb{I}_j^{\text{SAF}}} \right.\\
	\left. - \sum_{0}^{n_{\text{min}}}{\mathbb{I}_j^{\text{SAF}}  \rvert_{x(0)=x_{k_\text{ter}},I(t)=I_{\text{min}} }} \right),& \text{if $ k=k_{\text{ter}} $}\\\\
	0,& otherwise
	\end{cases}
	\end{eqnarray}
	where $ n_{\text{min}}=\left[\frac{3600(\text{SOC}_{\text{given}}-\text{SOC}_{k_{\text{ter}}})I_{\text{1Crate}}}{I_{\text{min}}t} \right]$.
	Reward function is composed of the SEI layer formation term and the safety constraint violation term to deal with the objective function (\ref{objective}) and the safety constraints (\ref{safety_1})-(\ref{safety_4}). To make the reward simple, there is no term to make the policy to satisfy final SOC condition (\ref{SOC_arrival_cond}). Instead, along with the termination condition, reward function is designed to always make the battery reach the desired SOC value $ \text{SOC}_{\text{given}} $ in the given time $ t_{\text{given}} $ with any policy of action without loss of generality.  
\end{itemize}
It is noted that $ \mathbf{s}_k $ and $ \mathbf{a}_k $ defined above are used as inputs of SAC networks ($ V_{\psi}(\cdot) $, $ V_{\bar{\psi}}(\cdot) $, $ Q_{\theta}(\cdot,\cdot) $, $ \pi(\cdot|\cdot) $) and $ \mathbf{s}_k $ is used to compute gradients of the objective in (\ref{Q_objective}) in this work.

\end{itemize}

\section{Result and discussion}
\begin{figure}[h!]\centering
	\includegraphics[width=0.5\textwidth]{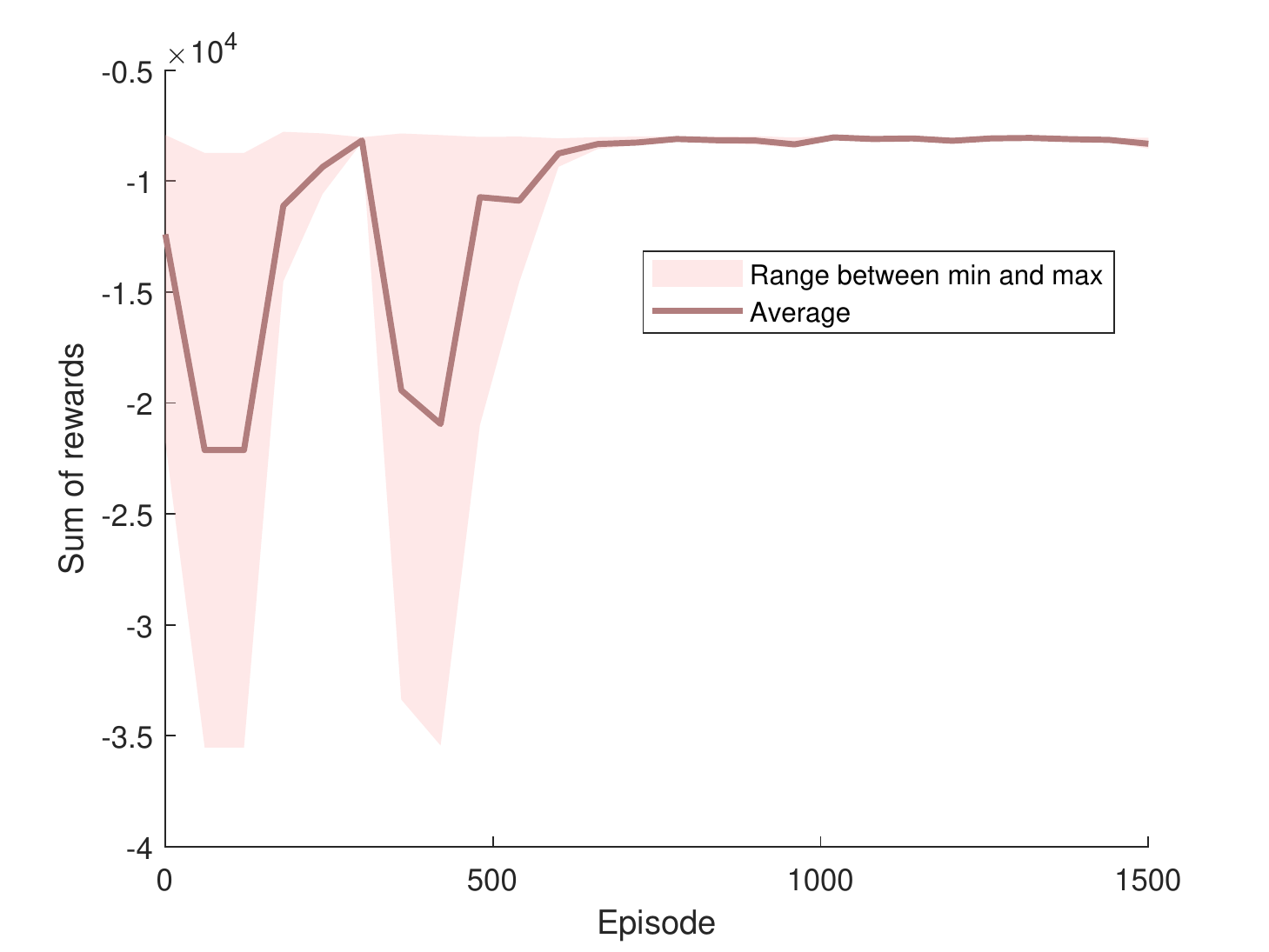}
	\caption{Change of reward sum throughout training episodes }\label{training_process}
\end{figure}

SAC (Section \ref{SAC}) is used to solve the optimal charging problem (Section \ref{optimal_charging_problem}) in this work. The following problem set and SAC neural networks are used to validate the effectiveness of the proposed optimal charging problem formulation and the proposed RL-based method to solve it:
\begin{itemize}[leftmargin=*]
	\item Optimal charging problem setting
		\begin{itemize}
			\item[-] $ \text{OCV}_{\text{SOC=0\%}}=3\ \text{V} $, $ \text{OCV}_{\text{SOC=100\%}}=4.2\ \text{V} $
			\item[-] $ \text{OCV}(t=0) = 3.3\ \text{V} $
			\item[-] $ \text{T}_{\text{jel}}(t=0), \text{T}_{\text{can}}(t=0) = 273.15 + 25\ \text{K} $
			\item[-] $ \text{SOC}_{\text{given}} = 0.8 $
			\item[-] $ I_{\text{min}} = 0$, $ I_{\text{max}} = 5 I_{\text{1Crate}} $
			\item[-] $ V_{\text{min}} = 2.8 \ \text{V}$, $ V_{\text{max}} = 4.5 \ \text{V} $
			\item[-] $ T_{\text{min}} = 273.15 + 0  \ \text{K}$, $ T_{\text{max}} = 273.15 + 45 \ \text{K} $
			\item[-] $ t_{\text{given}} \in [0.2\ \text{hr}, 2\ \text{hr}] $
		\end{itemize}
	
	\item SAC neural network setting
		\begin{itemize}
			
			\item[-] $ V_{\psi}(\cdot) $, $ V_{\bar{\psi}}(\cdot) $, $ Q_{\theta}(\cdot,\cdot) $, and $ \pi(\cdot|\cdot) $ are all composed of 4 layers of 256 neuron units with ReLU nonlinearity.
			
			\item[-] Discount factor  : 0.999
			\item[-] Learning rate of all networks : $ 1.0\times 10^{-4} $
			\item[-] Replay buffer size : $ 2.0\times 10^{6} $
			\item[-] Target smoothing coefficient : 0.005
			
		\end{itemize}
	\item Reward function parameter setting
		\begin{itemize}
			\item[-] $ \omega_{\text{SEI}} = 2.0\times 10^{10} $, $ \omega_{\text{SAF}} = 1.0\times 10^{2} $
		\end{itemize}
\end{itemize} 
In this validation, the battery is initially in the equilibrium state before charging. 
It is noted that $ t_{\text{given}} $ is given by a certain range of time to maximize the battery life effectively according to the various demands of the end-users. In the training phase, $ t_{\text{given}} $ is sampled from uniform distribution ($ t_{\text{given}} \sim U(0.2,2) $) at each training step. The figure \ref{training_process} shows how the sum of rewards converges to the maximum value during the training process. The $ x $-axis values represents the number of episodes used in the training. The $ y $-axis values are obtained by testing the trained SAC networks at every 60 episodes. In such test, 30 episodes are implemented for 30 different $ t_{\text{given}} $ which are sampled form uniform distribution with range $ [0.2 , 2] $ and then the average, minimum, and maximum values of reward sum is plotted.

As shown in the figure \ref{training_process}, SAC policy almost converges after about 700 episodes.

\begin{figure*}[htp]
	\centering
	\subfigure{\includegraphics[width=0.48\textwidth]{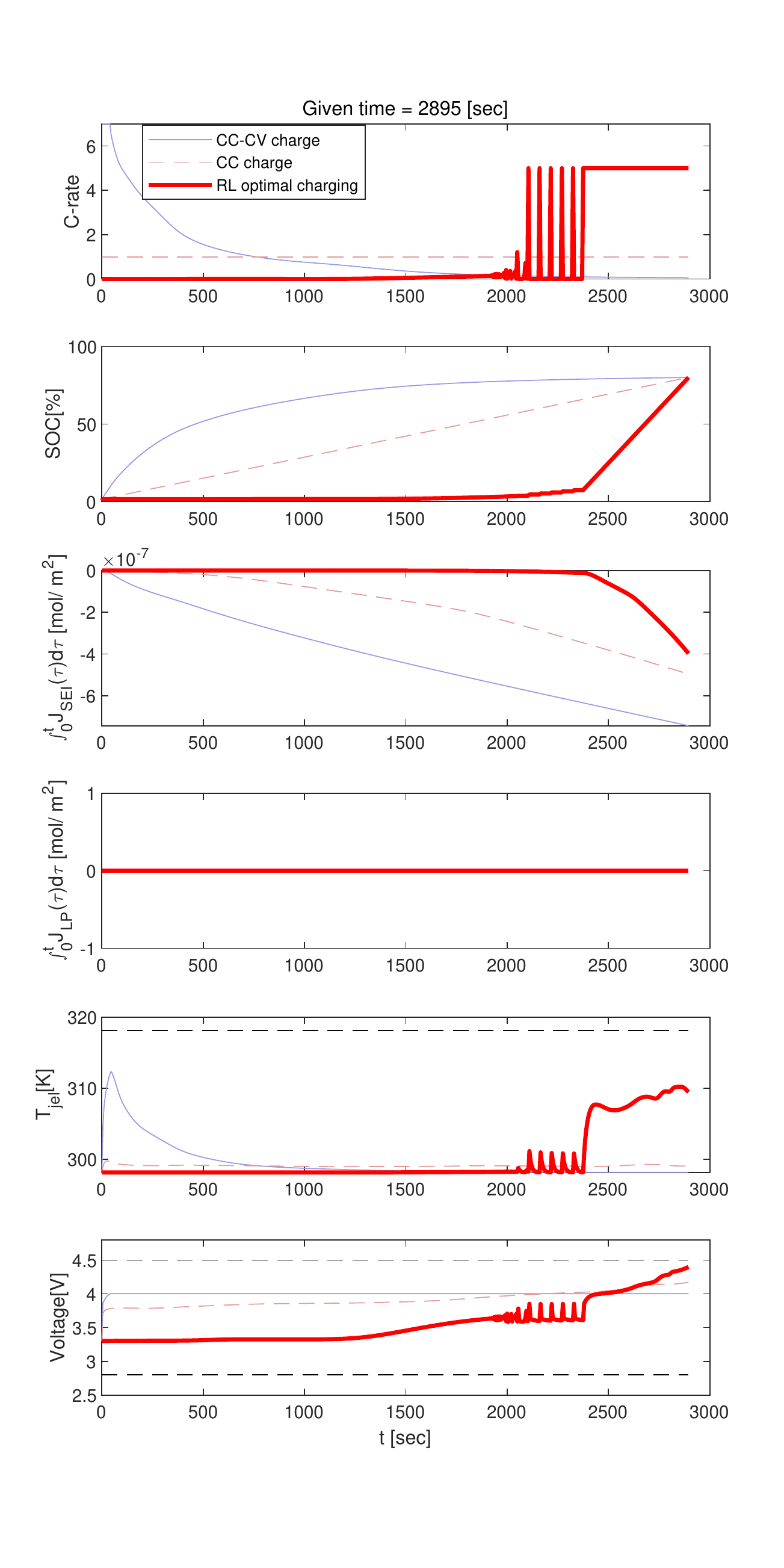}}\quad\quad
	\subfigure{\includegraphics[width=0.48\textwidth]{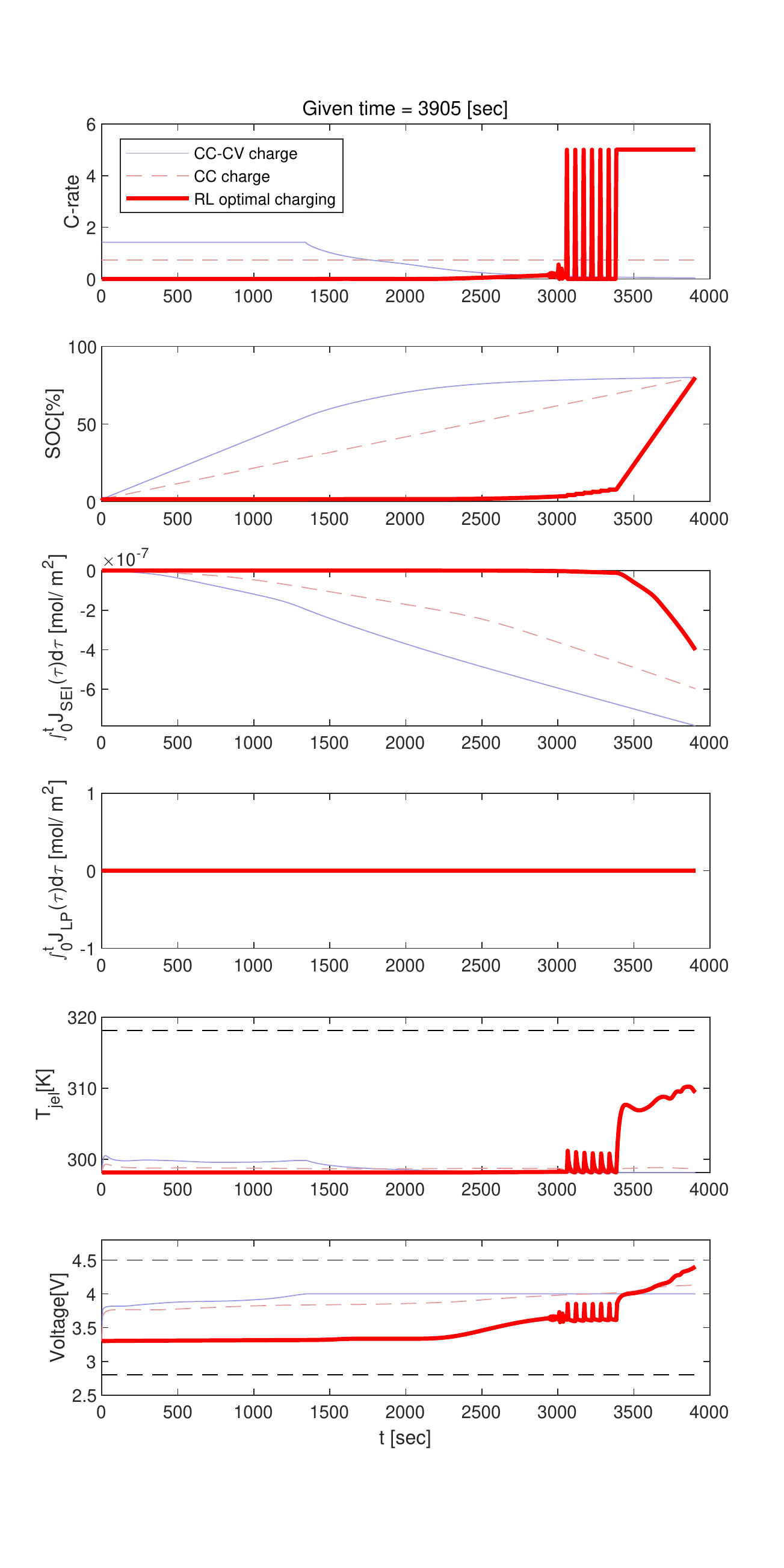}}
	\caption{Results of SAC-based optimal charging profile and benchmarks for different given charge times $ t_{\text{given}}  $ }
	\label{validation_results1}
\end{figure*}

\begin{figure*}[htp]
	\centering
	\subfigure{\includegraphics[width=0.48\textwidth]{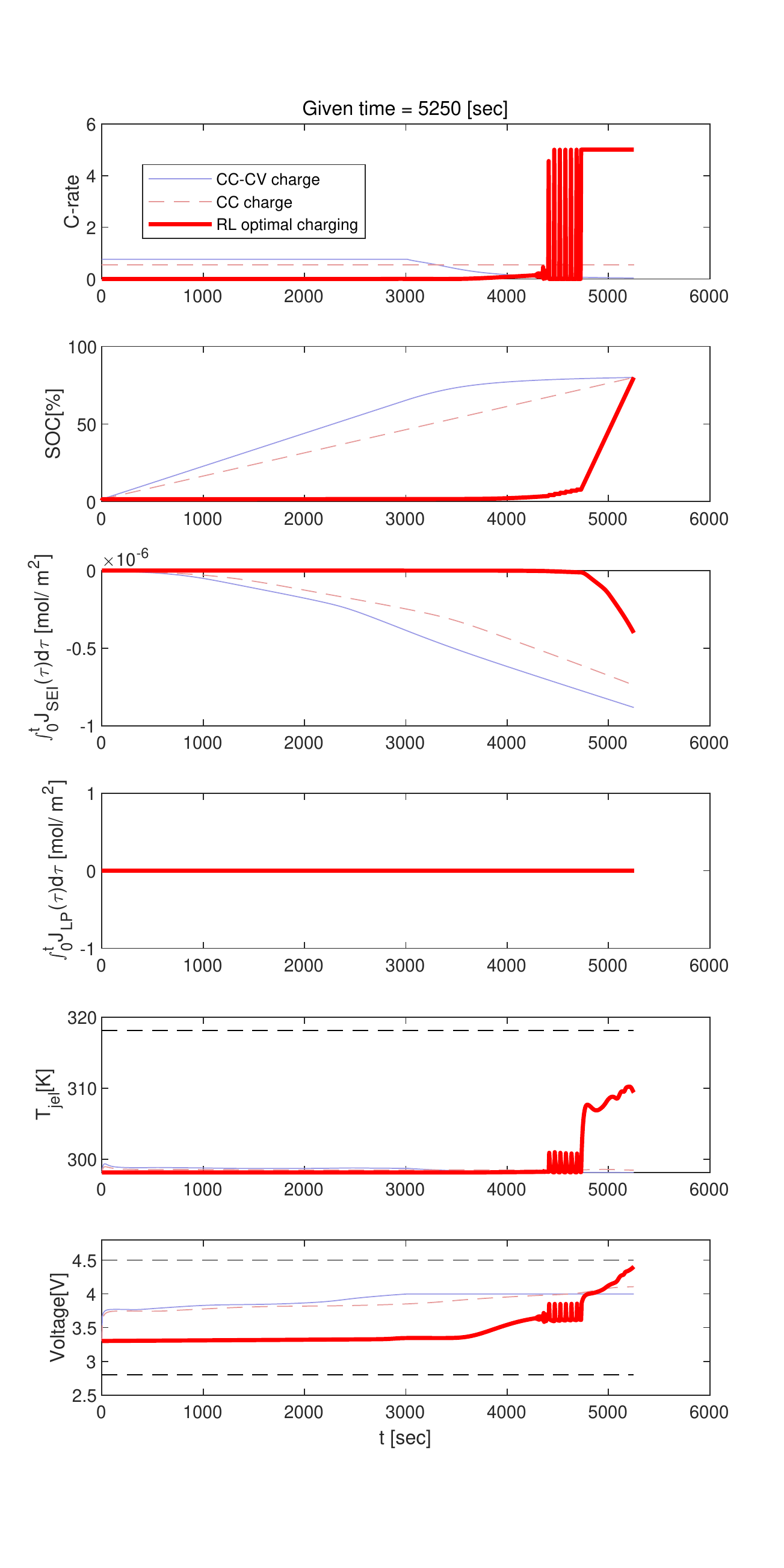}}\quad\quad
	\subfigure{\includegraphics[width=0.48\textwidth]{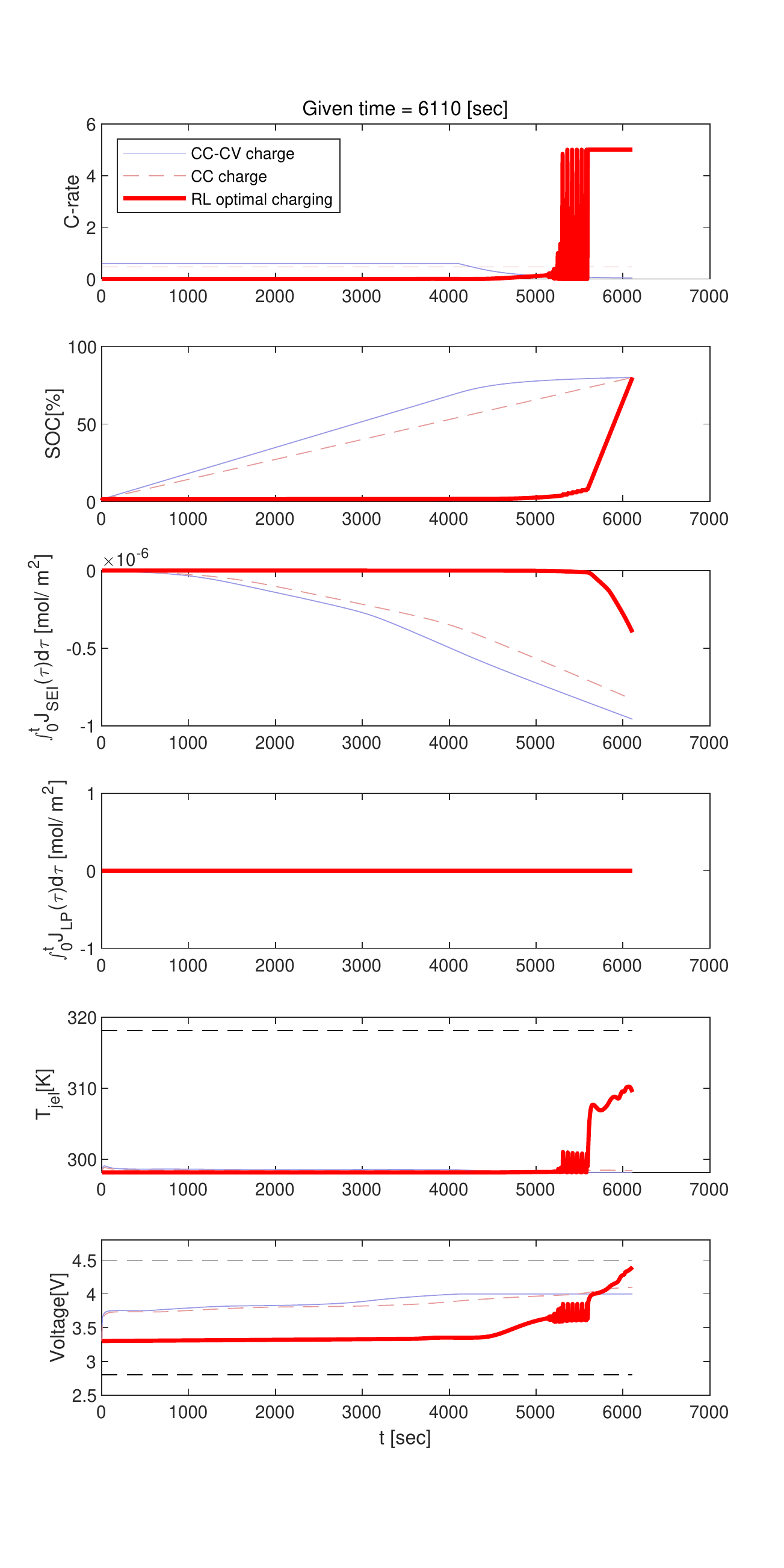}}
		\caption{Results of SAC-based optimal charging profile and benchmarks for different given charge times $ t_{\text{given}}  $ }
	\label{validation_results2}
\end{figure*}
The validation is implemented by comparing the result of SAC-based optimal charging profiles with those of constant current (CC) and CC-CV (constant voltage) charging profiles, which are widely used in many applications. It is noted that these three kinds of profiles have the same charge time, initial state, and goal SOC ($ \text{SOC}_{\text{given}} $) for the fair comparison of their performances.
As seen in the figure \ref{validation_results1} and figure \ref{validation_results2}, the proposed SAC-based optimal charging strategy has less amout of SEI layer formation during the charging process than any other benchmarks while not violating the constraints at the same time.

\begin{figure}[h!]\centering
	\includegraphics[width=0.5\textwidth]{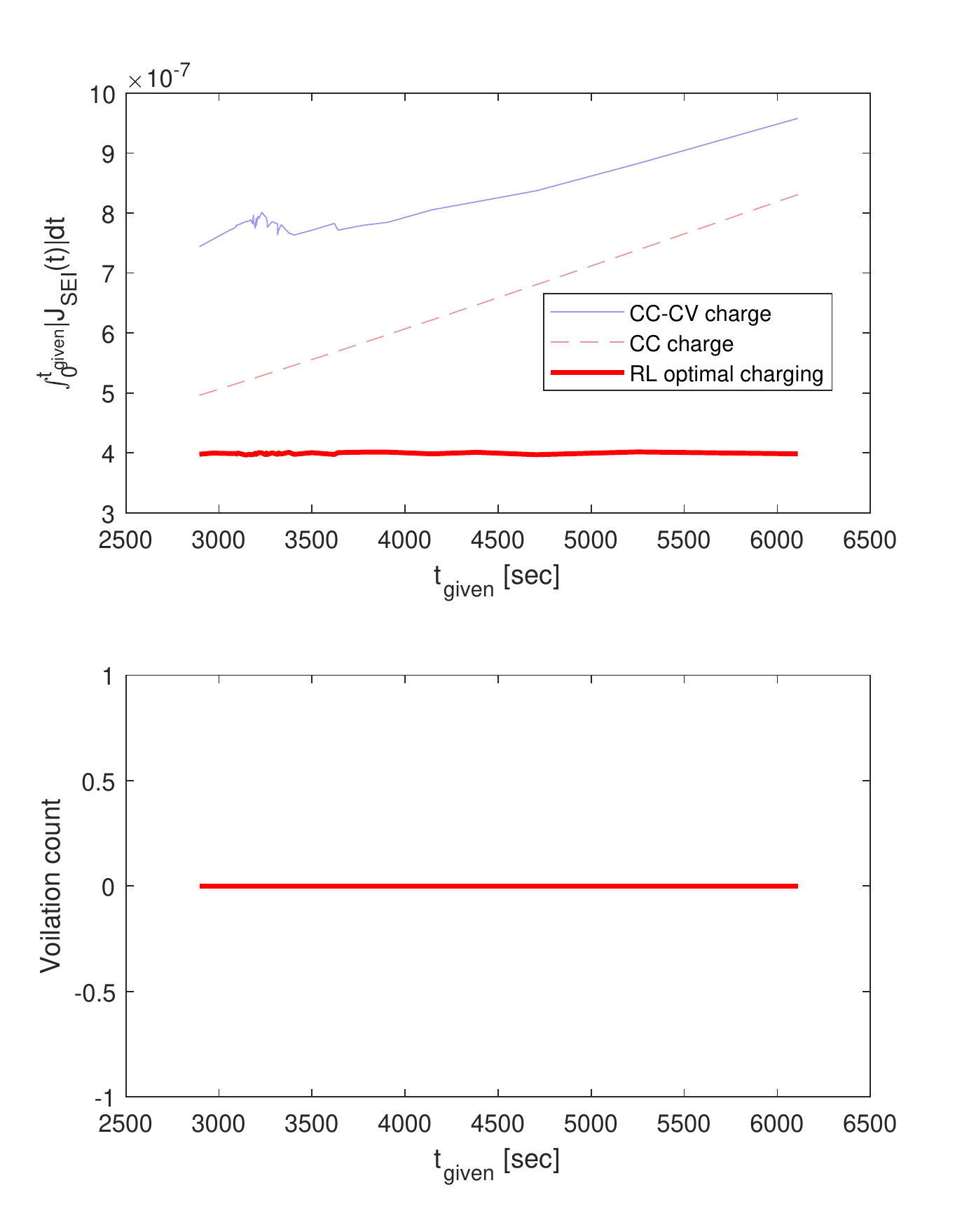}
	\caption{Amount of SEI layer formation and safety violation count of SAC-based optimal charging profiles and bench mark profiles. }\label{overall_result}
\end{figure}

Figure \ref{overall_result} is obtained by testing the SAC-based optimal charging and benchmarks for 40 different $ t_{\text{given}} $. The proposed SAC-based optimal charging strategy shows the best performance with respect to the SEI film formation and safety constraints than CC charge and CC-CV charge strategy.


\section{conclusion}
In this study, Optimal charging strategy for the life extension of the Li-ion battery using a reinforcement learning (RL) algorithms is proposed. Specifically, the optimal charging is defined to minimize SEI layer formation while various safety constraints related to voltage, current, temperature, and Li-plating. SPMeT-based battery aging model is designed and used in this study to make use of SPMeT's fast computation and high accuracy. Soft actor-critic (SAC), one of the state-of-the-art RL algorithms is employed to solve this optimal charging problem for its data-efficiency and stability of training process. Special termination condtions and reward function are proposed to improve the stability of the training process instead of reflecting all safety constraints in the reward function.  The proposed SAC-based optimal charging strategy is shown to have better performance than traditional charging strategy such as CC and CC-CV charging in the simulation of the proposed SPMeT-based battery aging model. Future studies will consider other complicated electrochemical phenomena inside the Li-ion battery and implement experimental validation for the proposed optimal charging strategy.



\normalsize

\begin{thebibliography}{18}
\providecommand{\natexlab}[1]{#1}
\providecommand{\url}[1]{\texttt{#1}}
\expandafter\ifx\csname urlstyle\endcsname\relax
  \providecommand{\doi}[1]{doi: #1}\else
  \providecommand{\doi}{doi: \begingroup \urlstyle{rm}\Url}\fi

\bibitem[Klein et~al.(2011)Klein, Chaturvedi, Christensen, Ahmed, Findeisen,
  and Kojic]{klein2011optimal}
Reinhardt Klein, Nalin~A Chaturvedi, Jake Christensen, Jasim Ahmed, Rolf
  Findeisen, and Aleksandar Kojic.
\newblock Optimal charging strategies in lithium-ion battery.
\newblock pages 382--387, 2011.

\bibitem[Torchio et~al.(2015)Torchio, Wolff, Raimondo, Magni, Krewer, Gopaluni,
  Paulson, and Braatz]{torchio2015real}
Marcello Torchio, Nicolas~A Wolff, Davide~M Raimondo, Lalo Magni, Ulrike
  Krewer, R~Bushan Gopaluni, Joel~A Paulson, and Richard~D Braatz.
\newblock Real-time model predictive control for the optimal charging of a
  lithium-ion battery.
\newblock pages 4536--4541, 2015.

\bibitem[Perez et~al.(2016)Perez, Hu, and Moura]{perez2016optimal}
Hector~Eduardo Perez, Xiaosong Hu, and Scott~J Moura.
\newblock Optimal charging of batteries via a single particle model with
  electrolyte and thermal dynamics.
\newblock pages 4000--4005, 2016.

\bibitem[Perez et~al.(2017)Perez, Hu, Dey, and Moura]{perez2017optimal}
Hector~Eduardo Perez, Xiaosong Hu, Satadru Dey, and Scott~J Moura.
\newblock Optimal charging of li-ion batteries with coupled
  electro-thermal-aging dynamics.
\newblock \emph{IEEE Transactions on Vehicular Technology}, 66\penalty0
  (9):\penalty0 7761--7770, 2017.

\bibitem[Zou et~al.(2017)Zou, Hu, Wei, Wik, and Egardt]{zou2017electrochemical}
Changfu Zou, Xiaosong Hu, Zhongbao Wei, Torsten Wik, and Bo~Egardt.
\newblock Electrochemical estimation and control for lithium-ion battery
  health-aware fast charging.
\newblock \emph{IEEE Transactions on Industrial Electronics}, 65\penalty0
  (8):\penalty0 6635--6645, 2017.

\bibitem[Yin et~al.(2019)Yin, Hu, Choe, Cho, and Joe]{yin2019new}
Yilin Yin, Yang Hu, Song-Yul Choe, Hana Cho, and Won~Tae Joe.
\newblock New fast charging method of lithium-ion batteries based on a reduced
  order electrochemical model considering side reaction.
\newblock \emph{Journal of Power Sources}, 423:\penalty0 367--379, 2019.

\bibitem[Park et~al.(2020)Park, Pozzi, Whitmeyer, Joe, Raimondo, and
  Moura]{park2020reinforcement}
Saehong Park, Andrea Pozzi, Michael Whitmeyer, Won~Tae Joe, Davide~M Raimondo,
  and Scott Moura.
\newblock Reinforcement learning-based fast charging control strategy for
  li-ion batteries.
\newblock \emph{arXiv preprint arXiv:2002.02060}, 2020.

\bibitem[Liu et~al.(2005)Liu, Teng, and Lin]{liu2005search}
Yi-Hwa Liu, Jen-Hao Teng, and Yu-Chung Lin.
\newblock Search for an optimal rapid charging pattern for lithium-ion
  batteries using ant colony system algorithm.
\newblock \emph{IEEE Transactions on Industrial Electronics}, 52\penalty0
  (5):\penalty0 1328--1336, 2005.

\bibitem[Suthar et~al.(2014)Suthar, Ramadesigan, De, Braatz, and
  Subramanian]{suthar2014optimal}
Bharatkumar Suthar, Venkatasailanathan Ramadesigan, Sumitava De, Richard~D
  Braatz, and Venkat~R Subramanian.
\newblock Optimal charging profiles for mechanically constrained lithium-ion
  batteries.
\newblock \emph{Physical Chemistry Chemical Physics}, 16\penalty0 (1):\penalty0
  277--287, 2014.

\bibitem[Bashash et~al.(2011)Bashash, Moura, Forman, and
  Fathy]{bashash2011plug}
Saeid Bashash, Scott~J Moura, Joel~C Forman, and Hosam~K Fathy.
\newblock Plug-in hybrid electric vehicle charge pattern optimization for
  energy cost and battery longevity.
\newblock \emph{Journal of power sources}, 196\penalty0 (1):\penalty0 541--549,
  2011.

\bibitem[Le~Floch et~al.(2016)Le~Floch, Belletti, and Moura]{le2016optimal}
Caroline Le~Floch, Francois Belletti, and Scott Moura.
\newblock Optimal charging of electric vehicles for load shaping: A
  dual-splitting framework with explicit convergence bounds.
\newblock \emph{IEEE Transactions on Transportation Electrification},
  2\penalty0 (2):\penalty0 190--199, 2016.

\bibitem[Pozzi et~al.(2018)Pozzi, Torchio, and Raimondo]{pozzi2018film}
Andrea Pozzi, Marcello Torchio, and Davide~M Raimondo.
\newblock Film growth minimization in a li-ion cell: a pseudo two dimensional
  model-based optimal charging approach.
\newblock In \emph{2018 European Control Conference (ECC)}, pages 1753--1758.
  IEEE, 2018.

\bibitem[Moura et~al.(2016)Moura, Argomedo, Klein, Mirtabatabaei, and
  Krstic]{moura2016battery}
Scott~J Moura, Federico~Bribiesca Argomedo, Reinhardt Klein, Anahita
  Mirtabatabaei, and Miroslav Krstic.
\newblock Battery state estimation for a single particle model with electrolyte
  dynamics.
\newblock \emph{IEEE Transactions on Control Systems Technology}, 25\penalty0
  (2):\penalty0 453--468, 2016.

\bibitem[Yang et~al.(2019)Yang, Hua, Qiao, Lian, Pan, and He]{yang2019coupled}
Shi-chun Yang, Yang Hua, Dan Qiao, Yu-bo Lian, Yu-wei Pan, and Yong-ling He.
\newblock A coupled electrochemical-thermal-mechanical degradation modelling
  approach for lifetime assessment of lithium-ion batteries.
\newblock \emph{Electrochimica Acta}, 326:\penalty0 134928, 2019.

\bibitem[Haarnoja et~al.(2018)Haarnoja, Zhou, Abbeel, and
  Levine]{haarnoja2018soft}
Tuomas Haarnoja, Aurick Zhou, Pieter Abbeel, and Sergey Levine.
\newblock Soft actor-critic: Off-policy maximum entropy deep reinforcement
  learning with a stochastic actor.
\newblock \emph{arXiv preprint arXiv:1801.01290}, 2018.

\bibitem[Andrychowicz et~al.(2017)Andrychowicz, Wolski, Ray, Schneider, Fong,
  Welinder, McGrew, Tobin, Abbeel, and Zaremba]{andrychowicz2017hindsight}
Marcin Andrychowicz, Filip Wolski, Alex Ray, Jonas Schneider, Rachel Fong,
  Peter Welinder, Bob McGrew, Josh Tobin, OpenAI~Pieter Abbeel, and Wojciech
  Zaremba.
\newblock Hindsight experience replay.
\newblock In \emph{Advances in neural information processing systems}, pages
  5048--5058, 2017.

\bibitem[Doyle et~al.(1993)Doyle, Fuller, and Newman]{doyle1993modeling}
Marc Doyle, Thomas~F Fuller, and John Newman.
\newblock Modeling of galvanostatic charge and discharge of the
  lithium/polymer/insertion cell.
\newblock \emph{Journal of the Electrochemical society}, 140\penalty0
  (6):\penalty0 1526, 1993.

\bibitem[Fuller et~al.(1994)Fuller, Doyle, and Newman]{fuller1994simulation}
Thomas~F Fuller, Marc Doyle, and John Newman.
\newblock Simulation and optimization of the dual lithium ion insertion cell.
\newblock \emph{Journal of the Electrochemical Society}, 141\penalty0
  (1):\penalty0 1--10, 1994.

\end{thebibliography}


\end{document}